\documentclass[aps,prl,reprint,superscriptaddress]{revtex4-1}
\usepackage{subfiles}
\usepackage{graphicx}
\usepackage{textcomp}
\usepackage{gensymb}
\usepackage{xfrac}
\usepackage{amsmath}
\usepackage{amssymb}
\usepackage{wrapfig}

\begin{document}

\title{Band Excitations in CePd$_3$: A Comparison of Neutron Scattering \\
and \textit{ab initio} Theory}

\author{Eugene A. Goremychkin}
\affiliation{Frank Laboratory of Neutron Physics, Joint Institute for Nuclear 
Research, Dubna, Moscow Region, 141980, Russia}
\author{Hyowon Park}
\affiliation{Materials Science Division, Argonne National Laboratory, Argonne, 
IL 60439-4845, USA}
\affiliation{Department of Physics, University of Illinois at Chicago, Chicago, 
IL 60607, USA}
\author{Raymond Osborn}
\email{rosborn@anl.gov}
\author{Stephan Rosenkranz}
\affiliation{Materials Science Division, Argonne National Laboratory, Argonne, 
IL 60439-4845, USA}
\author{John-Paul Castellan}
\affiliation{Institute for Solid State Physics, Karlsruhe Institute of Technology, 
D-76021 Karlsruhe, Germany}
\author{Victor R. Fanelli}
\affiliation{Instrument and Source Division, Oak Ridge National 
Laboratory, Oak Ridge, TN 37831, USA}
\author{Andrew D. Christianson}
\author{Matthew B. Stone}
\affiliation{Quantum Condensed Matter Division, Oak Ridge National 
Laboratory, Oak Ridge, TN 37831, USA}
\author{Eric D. Bauer}
\author{Kenneth J. McClellan}
\author{Darrin D. Byler}
\affiliation{Los Alamos National Laboratory, Los Alamos, NM 87545, USA}
\author{Jon M. Lawrence}
\affiliation{Los Alamos National Laboratory, Los Alamos, NM 87545, USA}
\affiliation{Department of Physics and Astronomy, University of California, 
Irvine, CA 92697, USA}

\date{\today}

\begin{abstract}
Intermediate valence compounds containing rare earth or actinide ions are
archetypal systems for the investigation of strong electron correlations.
Their effective electron masses of 10 to 50 times the free electron mass
result from a hybridization of the highly localized $f$-electrons with the
more itinerant $d$-electrons, which is strong enough that their properties
are dominated by on-site electron correlations. To a remarkable degree, this
can be modeled by the Anderson Impurity Model, even though the $f$-electrons
are situated on a periodic lattice. However, in recent years, there has been
increasing evidence that the dynamic magnetic susceptibility of intermediate
valence compounds is not purely local, but shows variations across the
Brillouin zone that have been ascribed to $f$-band coherence. So far, this
has been based on simplified qualitative models. In this article, we
present a quantitative comparison of inelastic neutron scattering from a
single crystal of CePd$_3$, measured in four-dimensional
(\textbf{Q},$\omega$)-space,  with \textit{ab initio} calculations, which are
in excellent agreement on an absolute scale. Our results establish that the 
\textbf{Q}-dependence of the scattering is caused by particle-hole excitations 
within $f$-$d$ hybridized bands that grow in coherence with decreasing 
temperature.
\end{abstract}

\maketitle

\section{Introduction}
The advent of pulsed neutron sources, which have an enhanced flux of high
energy epithermal neutrons, stimulated interest over thirty years ago in the
possibility of using inelastic neutron scattering to study single-electron
band structures~\cite{Sinha:1984vx, Cooke:1982eg}. The neutron cross section,
or scattering law, S$(\mathbf{Q},\omega)$, is proportional to the dynamic
magnetic susceptibility, $\chi^{\prime\prime}(\mathbf{Q},\omega)$, of the
band electrons~\cite{Balcar:1989us}. For non-interacting electrons, this is
derived from the familiar Lindhard susceptibility, whose imaginary part is
proportional to the joint density-of-states of the single-electron bands,
$\chi_0^{\prime\prime}(\mathbf{Q},\omega) \propto
\sum_\mathbf{k}{f_{\mathbf{k}+\mathbf{Q}}(1-f_{\mathbf{k}})\delta(E_{\mathbf{
k}+\mathbf{Q}}-E_{\mathbf{k}}-\omega)}$. The resulting scattering intensity
would be enhanced at momentum transfers, \textbf{Q}, and energy transfers,
$\omega$, that connect regions of high densities-of-state in the
single-electron bands, $E_\mathbf{k}$, whose states are occupied with
probability $f_\mathbf{k}$, so that neutrons could become a complementary
probe of the electronic structure to Angle Resolved Photoemission
Spectroscopy (ARPES).

\begin{figure*}[!ht]
\includegraphics[width =  2\columnwidth]{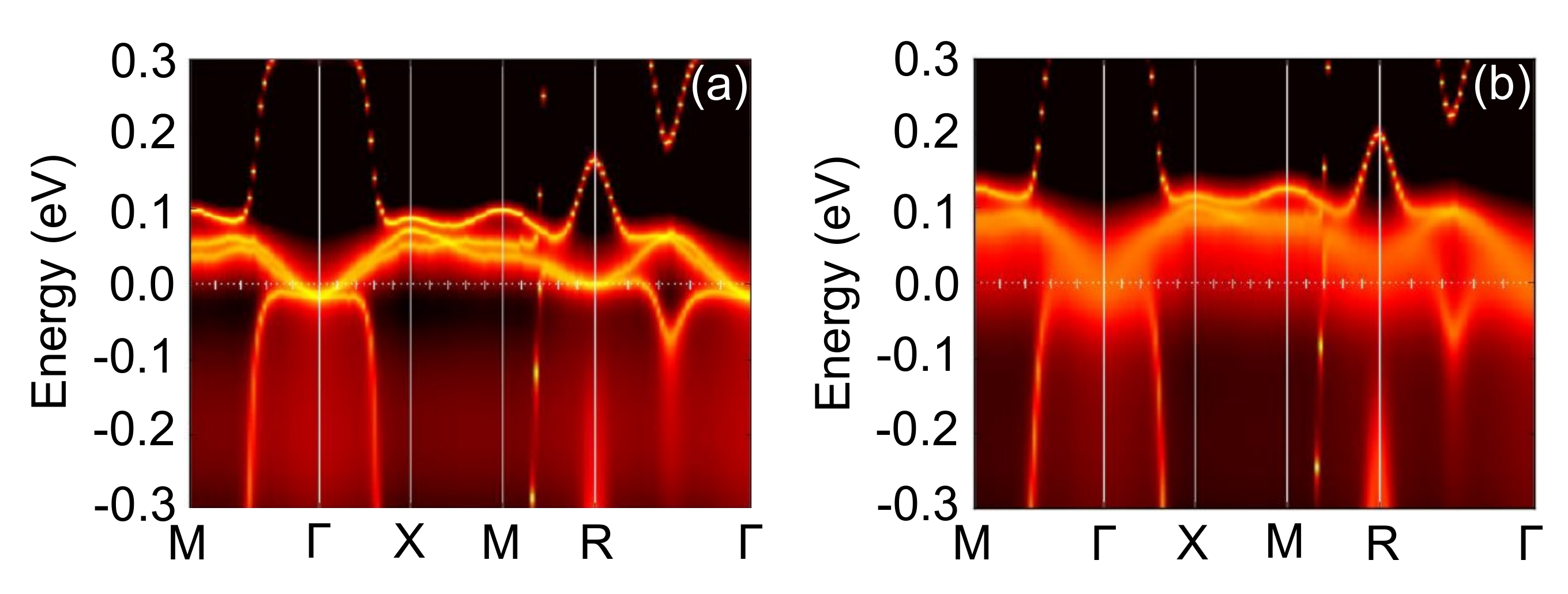}
\caption{Electronic spectral function of CePd$_3$ at (a) 100~K and (b) 400~K,
calculated by using DFT+DMFT. At low temperature, the calculations show the
presence of well-defined quasiparticle bands that cross the Fermi energy
producing small electron pockets close to the centre of the Brillouin zone
($\Gamma$) and the ($\frac{1}{2}\frac{1}{2}\frac{1}{2}$) zone boundary (R).
Flat unoccupied bands near 50 meV are seen at the X-point ($\frac{1}{2}$00)
and the M-point ($\frac{1}{2}\frac{1}{2}$0). At 400~K, the spectral weight close 
to the Fermi energy is largely incoherent.
\label{Figure1}}
\end{figure*}

The earliest estimates of the neutron cross section for weakly correlated
electron bands were discouraging, with signals in the range $10^{-4}$ to
$10^{-3}$ barns/steradians/eV spread over wave vectors covering the entire
Brillouin zone and energies up to the band width~\cite{Sinha:1984vx}. Such
broad distributions of intensity have been challenging to measure at the
pulsed neutron time-of-flight spectrometers, where measurements are typically
made using a fixed sample geometry. Consequently, such high-energy
spectrometers have mostly been used to measure coherent excitations, such as
spin waves in the copper oxide and iron-based
superconductors~\cite{Coldea:2001gu, Dai:2015eg}. These also represent the
magnetic excitations of band electrons, but they are easier to measure
because strong interatomic exchange interactions generate poles in the
dynamic susceptibility that yield well-defined peaks in the cross section.

Although measuring the electronic structure of weakly correlated electrons
may not be technically feasible with neutrons, there has been increasing
evidence over the past decade that strongly correlated electron systems, such
as the rare earth intermediate valence compounds, show variations in
$\chi^{\prime\prime}(\mathbf{Q},\omega)$ that could result from coherent
fermionic bands~\cite{Riseborough:2016gp}. Before this, it was generally
assumed that the strong correlations in these materials were purely local,
confined to $f$-electron interactions within each ion, so that the electronic
excitations of the $f$-shell were entirely
incoherent~\cite{HollandMoritz:1977ks}. However, downturns in the resistivity
at low temperature were interpreted as evidence of a growing coherence of
strongly renormalized $f$-bands whose electrons are hybridized with the more
itinerant $d$-electrons~\cite{Lawrence:1985fg}, an interpretation that was
reinforced by evidence of a \textbf{Q}-dependence of the scattering in single
crystals of materials such as CePd$_3$~\cite{Fanelli:2014bq},
YbAl$_3$~\cite{2006PhRvL..96k7206C} and CeInSn$_2$~\cite{Murani:2008jo}. So
far, these interpretations have been qualitative, based on conceptual models
of $f-d$ hybridized bands, because it was not possible to do a quantitative
comparison with realistic theories within the experimental limitations of
fixed-geometry neutron measurements.

Two recent developments allow us to go beyond the previous qualitative
analysis and do a comprehensive quantitative comparison of experiment and
theory. The first of these are theoretical advances in the calculation of
electronic excitations in strongly correlated electron systems, combining
Density Functional Theory (DFT) with Dynamical Mean Field Theory (DMFT), to
produce \textit{ab initio} calculations of the dynamic susceptibility that
include both single-particle and two-particle vertex corrections, allowing
the treatment of systems with both coherent quasiparticles and incoherent
spectral weight~\cite{Kotliar:2006fl}. We recently drew attention to the
qualitative consistency of our earlier neutron scattering data with strongly
renormalized band calculations in CePd$_3$~\cite{Fanelli:2014bq,
Sakai:2010kc}, but we have now performed a complete calculation of
S(\textbf{Q},$\omega$) over the entire Brillouin zone with cross sections
placed on an absolute scale. 

The second development is the advent of a new generation of inelastic neutron
scattering spectrometers with large position sensitive
detectors~\cite{Bewley:2006bl, Abernathy:2012hf} that allow efficient
measurements of four dimensional S(\textbf{Q},$\omega$) in single crystals by
rotating the sample during the data collection. This can be accomplished
either by measuring at discrete steps of the rotation angle, which is known
as the Horace mode, named after the analysis software developed by R. Ewings
and T. G. Perring~\cite{Ewings:2016ht}, or by collecting the data
continuously as the sample rotates, which is known as the Sweep
mode~\cite{Weber:2012iw}. Both methods produce equivalent results that
overcome the limitations of fixed-geometry measurements by measuring entire
volumes of (\textbf{Q},$\omega$)-space rather than a sparse set of
hypersurfaces through that volume. Since the experimental data can also be
placed on an absolute scale by normalizing the intensity to a vanadium
standard, it is possible to produce a parameter-free comparison of experiment
and theory.

In this article, we present the results of such a comparison between neutron
measurements obtained at the ISIS Pulsed Neutron Source and the Spallation
Neutron Source, collected by the Horace and Sweep modes, respectively, and
DFT+DMFT calculations. As expected, the calculations show broad distributions
of intensity with diffuse maxima at high symmetry points that shift within
the Brillouin zone as a function of energy transfer. These match the measured
distributions of the dynamic susceptibility with absolute cross sections that
are within 20\% of the theoretical predictions. Peaks in the dynamic
susceptibility at fixed momentum transfer are associated with values of
\textbf{Q} and $\omega$ that connect relatively flat regions of the coherent
quasiparticle bands, although incoherent scattering processes significantly
enhance the overall intensity. The consistency between theory and experiment
shows that the DFT+DMFT method provides a highly accurate prediction of the
electronic properties in materials with extremely strong electron
correlations.

\begin{figure*}[!ht]
\includegraphics[width =  2\columnwidth]{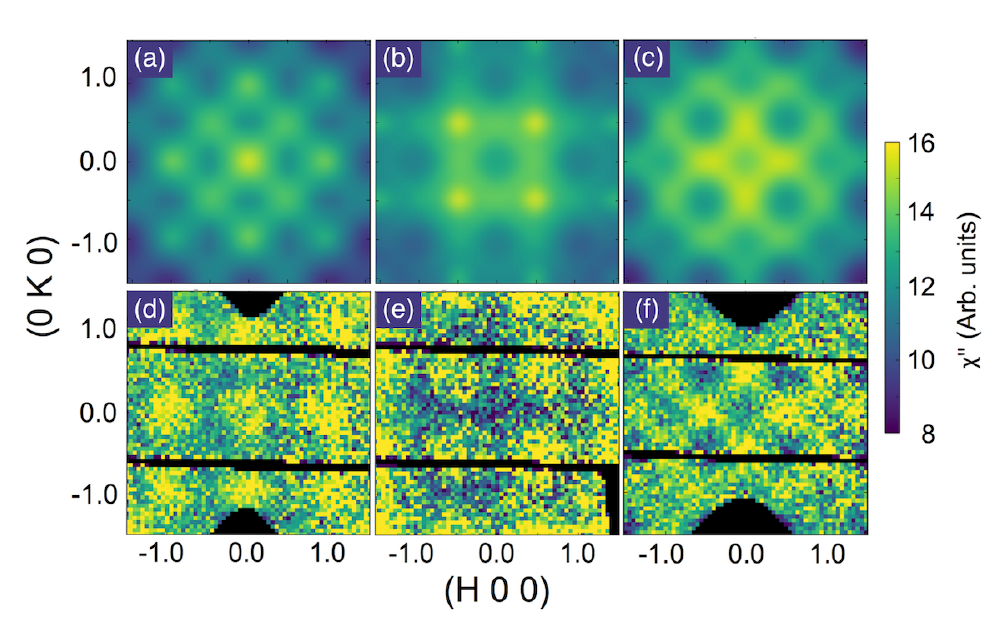}
\caption{Dynamic magnetic susceptibility of CePd$_3$ at 100~K (a,b,c) 
calculated by using DFT+DMFT and (d,e,f) measured by inelastic neutron 
scattering on ARCS. The results are shown as [H00]/[0K0] planes at 
constant energy transfers of (a,b,d,e) 35~meV and (c,f) 55~meV, with 
(a,d) L=1 and (b,c,e,f) L=$\frac{3}{2}$. In both the calculations and 
measurements, the results are averaged over $\omega=\pm5$~meV and 
$L\approx\pm0.2$. The intensity is in arbitrary units with the calculations and 
measurements normalized by a single scale factor. No backgrounds have been
subtracted from the ARCS data. Black pixels represent regions of reciprocal 
space that were not measured.
\label{Figure2}}
\end{figure*}

\section{Theory}

The dynamic susceptibility of CePd$_3$ was derived from a calculation of the
one-electron Green's function, using density functional theory in combination
with dynamical mean field theory (DFT+DMFT). This approach allows the
incorporation of local correlations, \textit{i.e.} Hund's rule and spin-orbit
coupling as well as on-site Kondo screening, into realistic band structures
based on DFT. Figure 1 shows the spectral function A(\textbf{k},$\omega$),
which includes the the one-particle vertex correction, \textit{i.e.}, the
electron self-energy. At 100~K, the band structure shows strongly
renormalized but well-defined quasiparticle excitations within $f$-electron
bands that are hybridized with the more dispersive $d$-bands. There are
two small Fermi surface pockets centred at the $\Gamma$, \textit{i.e.},
$\mathbf{Q}=$(000), and R-points, \textit{i.e.},
$\mathbf{Q}=(\frac{1}{2}\frac{1}{2}\frac{1}{2}$). Because of the strong
spin-orbit coupling, the $f$-bands have contributions from both
$j=\frac{5}{2}$ and $\frac{7}{2}$ states, with the former spread over
approximately 100~meV around the Fermi level and the latter at a few hundred
meV above the Fermi energy (not shown in Fig. 1). The spin-orbit coupling is
also evident in weak incoherent spectral weight at about 200~meV below the
Fermi energy. At 400~K, the overall dispersion of the quasiparticles is very
similar, but there is a strong increase in the self energy, with a strong
reduction of the coherent quasiparticle spectral weight, particularly close
to the Fermi energy. We will discuss this later when presenting the
high-temperature neutron scattering results.

The dynamic magnetic susceptibility,
$\chi^{\prime\prime}(\mathbf{Q},\omega)$, is computed from the
polarization bubble of the fully interacting DFT+DMFT one-particle Green's 
function, whose spectral weight is shown in Figure 1, by incorporating two-particle
irreducible vertex corrections, $\Gamma_{loc}^{irr}$, which are assumed to be 
local in the same basis in which the DMFT self-energy is 
local~\cite{Jarrell:1992gl,Park:2011fg}. Further details are given in the 
Supplementary Information.

\begin{figure}[!b]
\includegraphics[width =  \columnwidth]{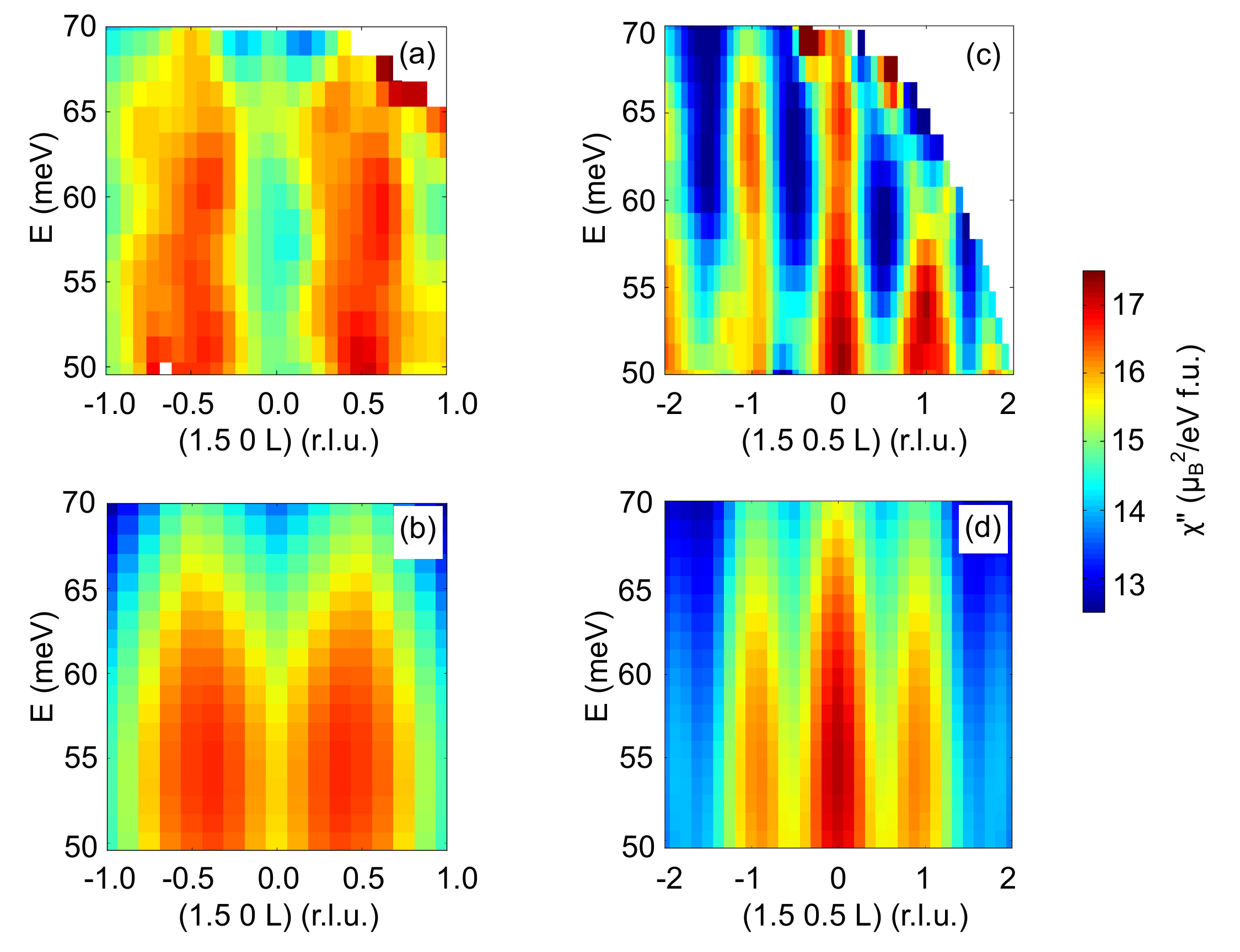}
\caption{Dynamic magnetic susceptibility of CePd$_3$ at 5 K ((a), (c) MERLIN 
data) and the DFT+DMFT calculations at 100~K (b), (d) represented by a 
L-$\omega$ slice at H = (1.5$\pm$0.25) and (a,b) K = (0.00$\pm$0.25) and 
and (c,d) K= (0.50$\pm$0.25). 
\label{Figure3}}
\end{figure}

The calculations generate scattering throughout the Brillouin zone with broad
maxima at high symmetry points in \textbf{Q}, which shift with energy
transfer. Figure 2 shows $\mathbf{Q}=[H,K]$ scattering planes with $L=0$
and $L=\frac{1}{2}$ at energy transfers of 35~meV and 55~meV. The
calculations are displayed in an extended zone scheme with corrections for
the $f$-electron magnetic form factor. These show that there are
maxima in the intensity at the $\Gamma$ and R-points at $\omega=35$~ meV, but
at the M and X points at $\omega=57$~meV. These maxima are not connected
by dispersive modes. Instead, the dynamic susceptibility is peaked at
$\sim35$~meV and $\sim57$~meV in different regions of the Brillouin zone.
This is illustrated in Figure 3, which shows slices in the $L-\omega$ plane
centred at the X and M-points. The scattering consists of columns of
intensity with maxima at the two discrete values of energy transfer.

\section{Experiment}
Using a large single crystal of CePd$_3$, we have performed measurements of
four-dimensional S(\textbf{Q},$\omega$) by rotating the sample at fixed
incident energies on time-of-flight spectrometers, MERLIN and ARCS, at the
ISIS Pulsed Neutron Facility and Spallation Neutron Source,
respectively~\cite{Bewley:2006bl, Abernathy:2012hf}. These possess large
banks of position-sensitive detectors that allow the scattered neutrons to be
counted as a function of polar and azimuthal angle, with respect to the
incident beam. When combined with the sample rotation angle and the neutron
time-of-flight, this four-coordinate scattering geometry can be readily
transformed into three-dimensional reciprocal space coordinates, \textbf{Q}, and 
a fourth energy coordinate, $\omega$. The transformed data fill large volumes of
(\textbf{Q},$\omega$), allowing arbitrary cuts to be made at constant energy
or momentum transfer (Fig. 2 and 3). Correction for the temperature factor and
calibration to a vanadium standard allows the dynamic magnetic susceptibility to 
be directly compared to the DFT+DMFT calculations. Uncertainties in the
absorption correction mean that there is 20\% uncertainty in the cross
section.

\begin{figure}[!t]
\includegraphics[width =  \columnwidth]{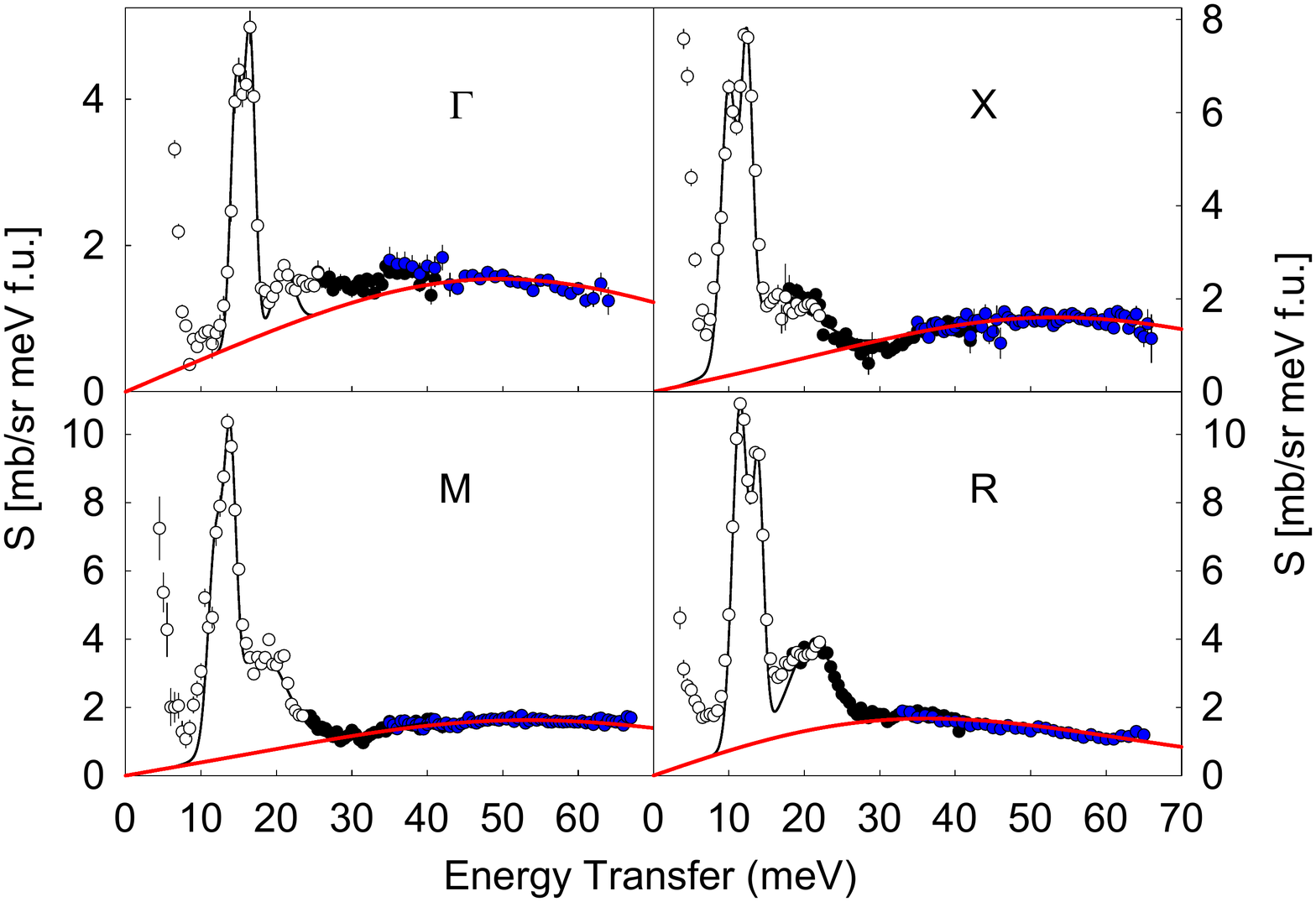}
\caption{Energy dependence of the scattering law of CePd$_3$ at 5~K at the 
$\Gamma$ (1.00$\pm$0.25, 1.00$\pm$0.25, 0.00$\pm$0.25), X (1.00$\pm$0.25, 
1.00$\pm$0.25, 0.50$\pm$0.25), M (1.50$\pm$0.25, 0.00$\pm$0.25, 
0.50$\pm$0.25) and R (1.50$\pm$0.25, 0.50$\pm$0.25, 0.50$\pm$0.25) points 
in the Brillouin zone. The open circles, closed black circles, and closed blue 
circles are measurements on MERLIN with incident energies of 30~meV, 
60~meV, and 
120~meV, respectively. The solid black lines show the estimated phonon 
scattering based on a fit to three Gaussians. The energies of the phonon 
peaks are in a good agreement with previous measurements of the CePd$_3$ 
phonons~\cite{Loong:1988ir}. The red lines are the result of the DFT+DMFT 
calculations.
\label{Figure4}}
\end{figure}

\begin{figure*}[!t]
\includegraphics[width =  1.6\columnwidth]{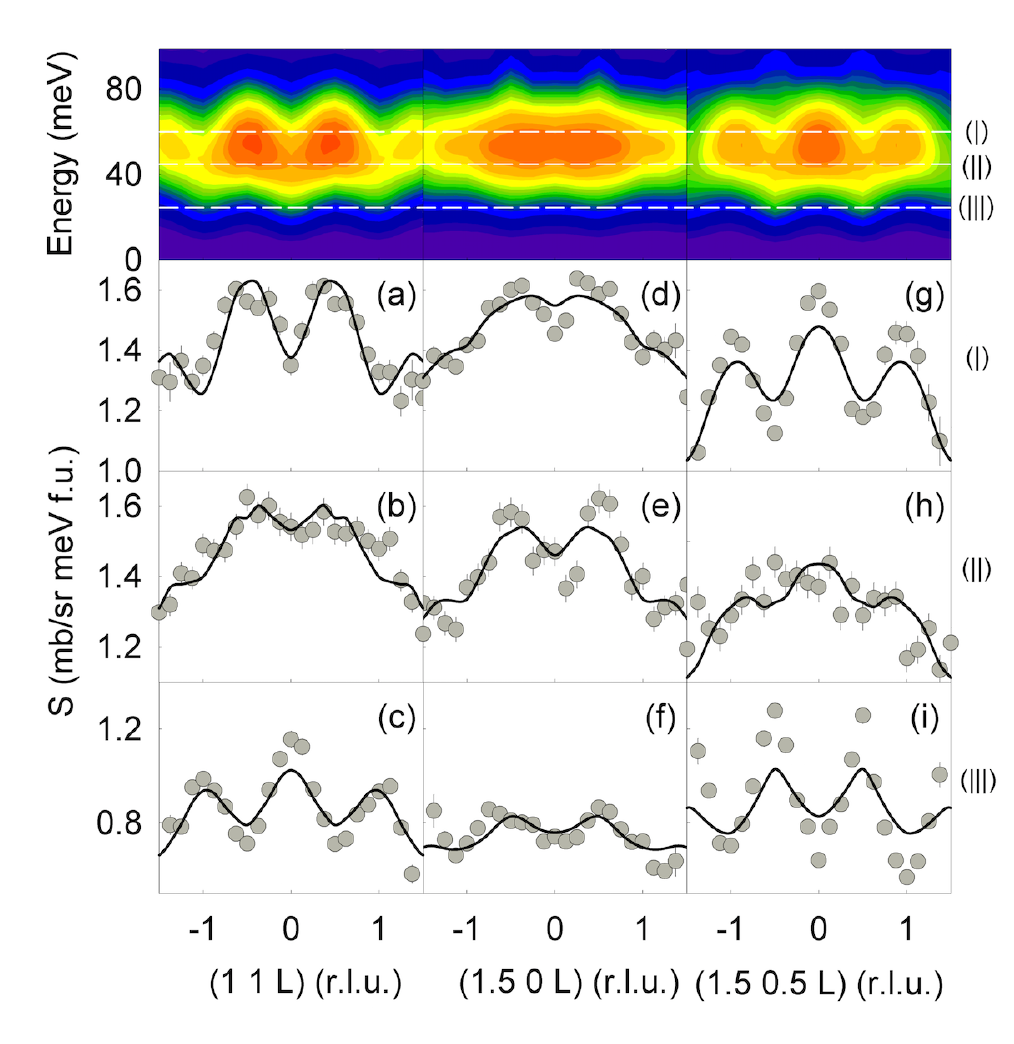}
\caption{The upper three panels represent the DFT+DMFT calculations of
S$(\mathbf{Q},\omega)$ at 100~K in the form of a L-$\omega$ slice along
$\Gamma$$\rightarrow$X at H=(1.00$\pm$0.25), K=(1.00$\pm$0.25),
X$\rightarrow$M at H=(1.50$\pm$0.25), K=(0.00$\pm$0.25) and M$\rightarrow$R
directions at H=(1.50$\pm$0.25), K=(0.50$\pm$0.25). Three white dashed lines,
labelled I, II and III, show the direction and position in energy of the one
dimensional constant energy cuts at energies of (a,d,g) 60$\pm$5~meV (b,e,h)
45.0$\pm$2.5~meV and (c,f,i) 25.0$\pm$2.5~meV, respectively. The points
represent the inelastic neutron scattering data measured on MERLIN and
integrated in the same energy and H, K range as in the upper three panels.
The error bars are derived using Poisson statistics. The lines represent
one-dimensional cuts of the DFT+DMFT calculations, which have been fit to the
data with a single scale factor and quadratic backgrounds as the only
adjustable parameters. In the plots, the backgrounds have been subtracted but
the data and fits without the background subtraction are shown in the
Supplementary Information.
\label{Figure5}}
\end{figure*}

Figure 4 shows the measured and calculated energy dependence of the
scattering at four points in the Brillouin zone. The excellent agreement
between the two without adjusting any of the theoretical parameters shows
that the DFT+DMFT calculations accurately reproduces the energy scale for the
magnetic fluctuations. The shift in the energy maximum from 35~meV at the
$\Gamma$ and R-points to 57~meV at the M and X-points is evident in the data.
The energy dependence of the scattering is also compared to the calculations
in Figure 3.

Figure 2 shows the \textbf{Q}-dependence derived from experiment, confirming 
the theoretically predicted shift in the maxima between the $\Gamma$ and 
R-points at 35 meV to the M and X-points at 57 meV. The magnetic scattering is 
superposed on a Q-dependent background from the sample environment, which 
increases monotonically with momentum transfer, shifting the maxima away 
from $Q=0$.

To provide a more quantitative comparison, we show constant energy cuts
along a number of high symmetry directions in Figure 5, where the data are
plotted against the theoretical calculations on an absolute scale. The
instrumental background, which is produced by scattering off the sample
environment, is well-described by a quadratic function in Q, so the data have
been fitted to the calculated values added to this background. Small
adjustments to the overall scale factor of the calculated dynamic
susceptibility were included in the fits, but these amounted to less than
20\%, which is consistent with the absorption correction from the irregularly
shaped sample. Details of these fits are given in the Supplementary Information. 
The excellent agreement confirms the qualitative consistency of experiment and 
theory evident in Figures 2 and 3.

\begin{figure}[!t]
\includegraphics[width =  \columnwidth]{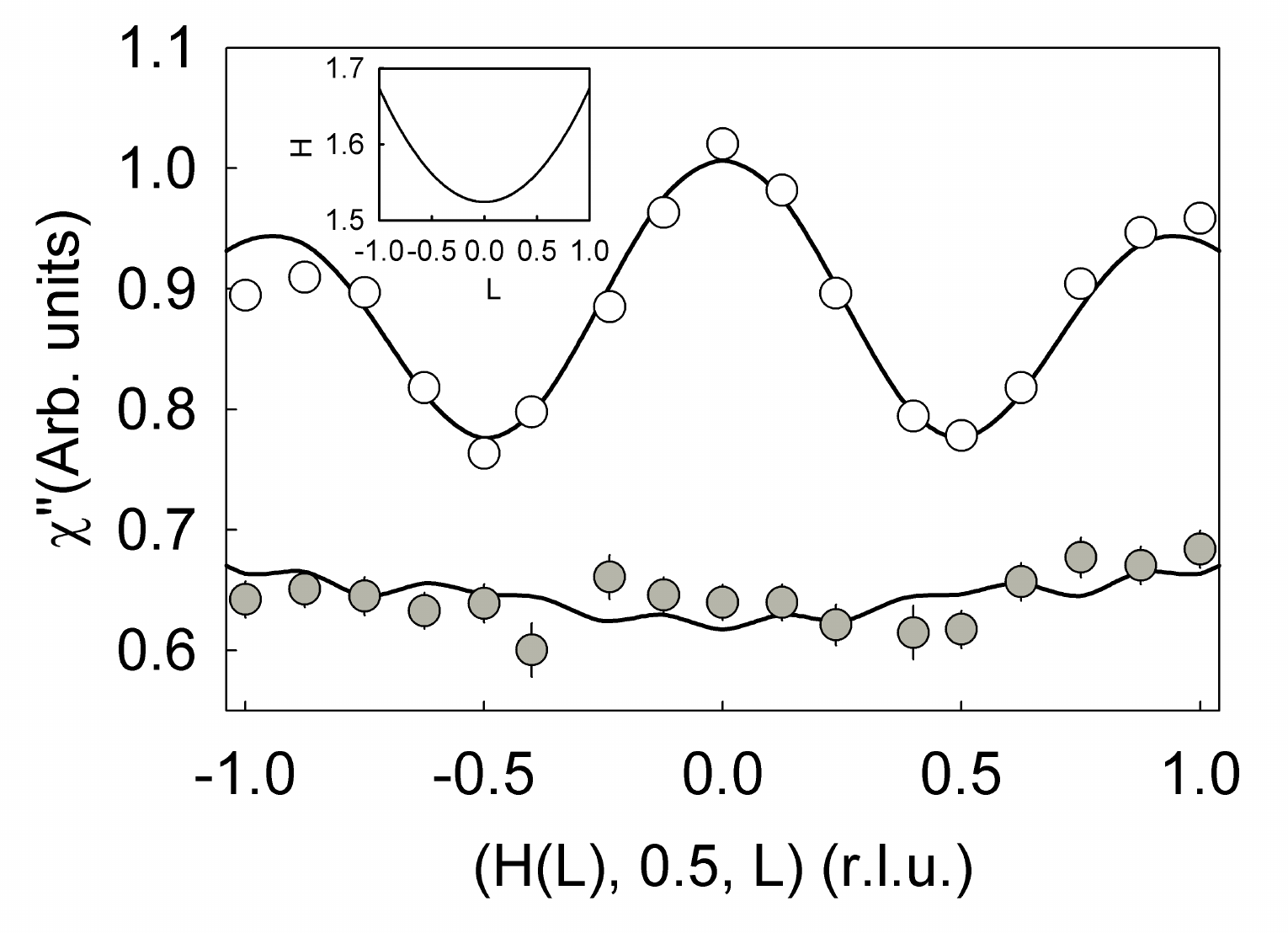}
\caption{Temperature dependence of the dynamic magnetic susceptibility
measured on the MAPS spectrometer. The open (closed) circles were measured at
6~K (300~K) using a fixed sample geometry (See Ref.~\citenum{Fanelli:2014bq}
for more details) at the energy transfer of (60$\pm$10) meV and K =
(0.50$\pm$0.25). The inset shows how H varies as a function of L. The lines
are the results of DFT+DMFT calculations performed at 100~K and 400~K
including the H(L) dependence and scaled with the same normalization factor
for both temperatures.
\label{Figure6}}
\end{figure}

Finally, we compare the calculations made at higher temperature to the
experimental data. Figure 6 shows that the Q-dependence of the magnetic
scattering is almost entirely suppressed at room temperature, which is well
above the coherence temperature inferred from transport measurements. This is
also predicted by the theoretical calculations. An inspection of the spectral
functions at 100~K and 400~K in Figure 1 shows that this results from the
substantial reduction in coherence of the one-electron quasiparticles, and
confirms the importance of this coherence in generating the observed
Q-variations in the dynamic susceptibility at low temperature.

\section{Discussion and Conclusion\label{conclusion}}
Our results demonstrate that it is now possible to determine the electronic
structure of intermediate valence materials with considerable accuracy by
incorporating local correlations into band structures through the combination
of density functional theory and dynamical mean field theory. The agreement
of the calculated and measured neutron cross sections is complete throughout
the Brillouin zone and extending over a broad range of energy transfers. This
also shows that inelastic neutron scattering can be used to measure the dynamic
susceptibility arising from correlated electron bands without the presence of
collective excitations, such as spin waves or crystal field excitons.

In the Supplementary Information, we compare the calculated magnetic response
both with and without the two-particle vertex correction,
$\Gamma_{loc}^{irr}$, which represents the interactions between the electron
and the hole excited by the neutron. The correction has two effects: firstly,
it smooths out some of the fine structure in the energy dependence of the
spectra while broadly preserving both the \textbf{Q}-variation and the
overall energy scale, and secondly, it produces a strong energy-dependent
enhancement of the intensity, for example by a factor of $\sim$6 at
$\omega=60$~meV (Fig. S4 in the Supplementary Information). Both coherent and
incoherent processes therefore make important contributions to the
scattering; the \textbf{Q}-dependence is determined primarily by the former
whereas the energy dependence and the overall intensity are determined
primarily by the latter.

In Ref.~\citenum{Fanelli:2014bq}, we showed that the Anderson
\textit{Impurity} Model (AIM) is  successful in explaining a number of
important properties (the magnetic susceptibility $\chi (T)$, the
4$f$-occupation number $n_f (T)$, the 4$f$ contribution to the specific heat
$C_{4f} (T)$, and the  \textbf{Q}-averaged dynamic susceptibility $\chi
^{\prime\prime} (\omega)$) of intermediate valence compounds such as
CePd$_3$, even though the cerium atoms are not impurities but sit on a
periodic lattice. We speculated that the reason the impurity model works so
well for these periodic systems is that the strong inelastic scattering of
the electronic quasiparticles broadens the spectral functions. The first
three properties just mentioned are primarily sensitive to the fourth,
\textit{i.e.}, $\chi ^{\prime\prime} (\omega)$, which represents
\textit{local} 4$f$ moment fluctuations. The DFT+DMFT calculations show that
the vertex corrections do indeed result in spectra that, when averaged in \textbf{Q},
are very similar to the AIM result of Ref. 9, thus helping explain why the
impurity model works as well as it does. These inelastic processes are also
responsible for the very rapid loss of coherence with temperature; as shown
in Figure 6 and in Ref. 9, the spectrum of CePd$_3$ is nearly
\textbf{Q}-independent at room temperature, and has the quasielastic spectral
shape expected for an Anderson impurity.

The results of this comparison between theory and experiment provide
significantly new insight into the nature of the correlations in intermediate
valence systems. The magnetic fluctuations show a much richer structure than
was implied by earlier toy models of hybridized bands, and there is a complex
interplay between coherent and incoherent contributions to the electronic
spectra that is reflected in the evolution of the dynamic magnetic susceptibility 
with temperature. The transition from coherent $f$-electron bands to local 
moment physics, so long postulated in models of heavy fermion and 
intermediate valence systems, is seen to be accurately modeled by the latest 
advances in the \textit{ab initio} theories of correlated electron systems.

\section{Methods\label{Methods}}

\subsection{Synthesis}
The single crystal of CePd$_3$ was grown using a modified Czochralski 
method. A polycrystalline charge was melted on a water-cooled, copper 
hearth in a tri-arc furnace under an argon atmosphere. The cooled hearth 
kept the molten charge contained within a solid skin of itself to avoid 
contamination of the melt.  A seed crystal mounted on a water-cooled 
rod was dipped into the molten compound and pulled upwards while being 
rotated. By controlling the temperature, the rate of pulling, and the speed of 
rotation, a single-crystalline cylindrical ingot of 0.5 cm diameter and 5 cm long,
with a mass of 17.72 g was produced. The sample was aligned with a [1,0,0]
direction held vertical.

\subsection{Inelastic Neutron Scattering}
Inelastic neutron scattering measurements were carried out using the MERLIN
and MAPS spectrometers at the ISIS neutron scattering facility and the ARCS
spectrometer at the Spallation Neutron Source. The measurements on MERLIN
were performed at 5~K using incident energies of 30~meV, 60~meV and 120~meV.
In these measurements, the crystal was mounted with [100] direction vertical
and angle $\phi$ between the incident wavevector, $k_i$ and the [010]
direction was rotated in discrete steps of $2^\circ$ (Horace mode). The data
were analyzed using the Horace software application~\cite{Ewings:2016ht}. The
ARCS measurements were performed using an incident energy of 120~meV using
the so-called Sweep mode, where $\phi$ was rotated continuously and neutron
counts were collected as time-stamped events that were synchronized with the
sample rotation angle. The data were analyzed using the Mantid
framework~\cite{Arnold:2014iy}. The MAPS experiment was performed at fixed
$\phi=0$, \textit{i.e.}, $k_i\|[010]$ at 7~K and 300~K.

\subsection{Theory}
We performed the charge self-consistent DFT+DMFT calculation and
computed the magnetic susceptibility of CePd$_3$ using the Wien2k+DMFT
package~\cite{Haule:2010dd}. The DFT equation was solved using the full
potential linearized augmented plane-wave method with the
Perdue-Burke-Ernzerhof exchange-correlation functional as implemented in the
Wien2k code~\cite{Wien2K}. The quantum impurity problem within DMFT was 
solved using the continuous-time quantum Monte Carlo (CTQMC)
method~\cite{Haule:2007kx, Werner:2006ko}. We set the on-site Coulomb
interaction $U=6$~eV and the Hund's coupling $J=0.7$~eV, as previously used for
elemental Ce in the $\alpha$ and $\gamma$ phases~\cite{Chakrabarti:2014vq}.
We set the lattice constant of the CePd$_3$ cubic unit cell to 4.126\AA, as 
measured in experiment. More details are given in the Supplementary Information.

\section{End Notes}
\subsection{Acknowledgements}
The research at the Joint Institute for Nuclear Research was supported by the
Russian Foundation for Basic Research project 16-02-01086. The research at
Argonne National Laboratory and Los Alamos National Laboratory was supported
by the Materials Sciences and Engineering Division, Office of Basic Energy
Sciences, U.S. Department of Energy. The research at Oak Ridge National
Laboratory's Spallation Neutron Source was supported by the Scientific User
Facilities Division, Office of Basic Energy Sciences, U.S. Department of
Energy.  Neutron experiments were performed at the Spallation Neutron Source,
Oak Ridge National Laboratory, USA, and the ISIS Pulsed Neutron Source,
Rutherford Appleton Laboratory, UK. We gratefully acknowledge the computing 
resources provided on Blues, a high-performance computing cluster operated by 
the Laboratory Computing Resource Center at Argonne National Laboratory. We 
are also grateful for useful discussions with Peter Riseborough.

\subsection{Author contributions} 
The single crystal was grown by E.D.B., K.J.M, and D.D.B. The experiments
were devised by J.M.L., E.A.G., V.R.F., A.D.C., S.R, and R.O. The inelastic
neutron scattering experiments  were performed by J.M.L., E.A.G., V.R.F.,
A.D.C., J.P.C., S.R, R.O, and. M.B.S. The data were analyzed by E.A.G, J.P.C., 
S.R., V.R.F. and J.M.L. The theoretical calculations were performed by H.P. The 
manuscript and supplementary information were written by R.O., E.A.G., J.M.L., 
and H.P. with input from all the authors.

\begin{figure}[!h]
\includegraphics[width=\columnwidth]{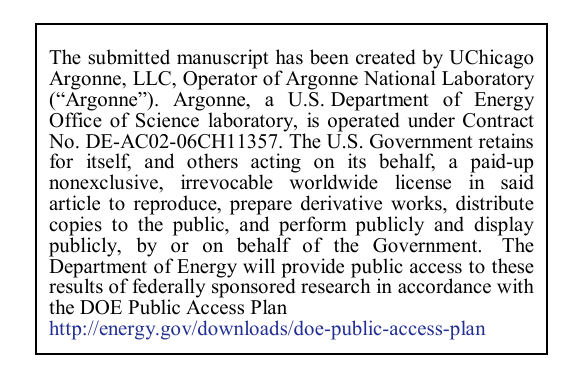}
\end{figure}
\newpage

\title{Supplementary Information}

\maketitle

\section{Theory}
\subsection{DFT+DMFT Method}
The \textit{ab initio} calculations of the dynamic magnetic susceptibility 
combine realistic band structures, calculated using density function theory 
(DFT), followed by solving the quantum impurity problem within Dynamic Mean
Field Theory (DMFT), in order to incorporate strong on-site correlations into 
the one-particle Green function. This DFT+DMFT approach has proved to be a 
powerful way of obtaining one-particle spectra that include both coherent 
quasiparticles and incoherent spectral weight from self-energy 
corrections~\cite{SKotliar:2006fl}.

The localized Ce $4f$ orbitals are constructed from a projector function in
such a way that the $f$ character is maximized, with self-consistent DFT
calculations performed using the Wien2k+DMFT package~\cite{SHaule:2010dd}. The
DFT+DMFT equation is solved using the full potential linearized augmented
plane-wave method with the Perdue-Burke-Ernzerhof exchange-correlation
functional as implemented in the Wien2k code~\cite{SWien2K}. The quantum
impurity problem within DMFT is solved using the continuous-time quantum
Monte Carlo (CTQMC) method~\cite{SHaule:2007kx,SWerner:2006ko}. We set the
on-site Coulomb interaction $U$=6eV and the Hund's coupling $J$=0.7eV as
previously used for the elemental Ce in $\alpha$ and $\gamma$
phases~\cite{SChakrabarti:2014vq}. The lattice constant of the CePd$_3$ cubic
unit cell used in the calculation is 4.126\AA$\:$ as measured in experiment.
The calculated one-particle spectral weight is shown in Fig. S1.

\begin{figure}[!h]
\includegraphics[width =  0.8\columnwidth]{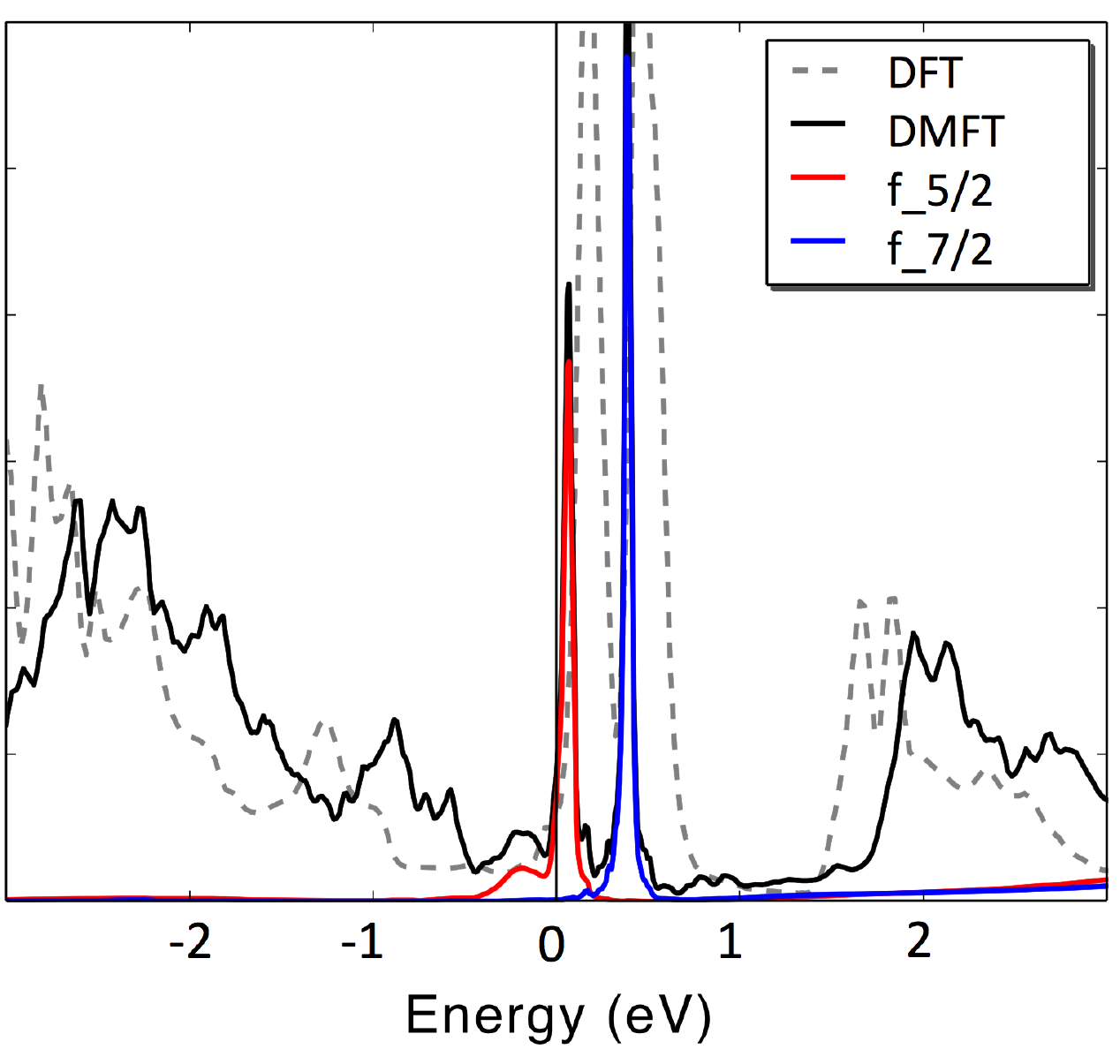}
\caption{Spectral weight of CePd$_3$ of the DFT and the DFT+DMFT 
calculations integrated over the Brillouin zone, showing a decomposition of
the $f$-electron spectra into contributions from different spin-orbit states
with $j=\frac{5}{2}$ (red) and $j=\frac{7}{2}$ (blue). 
\label{FigureS1}}
\vspace{-0.2in}
\end{figure}

The neutron scattering cross section is proportional to the imaginary part of the 
magnetic susceptibility, $\chi(\mathbf{q},\omega)$, which is computed using a
two-particle vertex method as developed in Ref.~\citenum{Park:2011fg}.
$\chi(\mathbf{q},i\omega_n)$ on the imaginary Matsubara frequency $i\omega_n$
axis can be obtained by summing the two-particle Green's function
$\chi(i\nu,i\nu^{\prime})_{\textbf{q},i\omega}$ closed by the total magnetic
moment vertex ($\mathbf{\mu}_J=g_J\mu_B\mathbf{J}$) over fermionic Matsubara
frequencies ($i\nu,i\nu^{\prime}$) and magnetic moment indices
($\alpha_{1-4}$):
\begin{equation}
\chi(\textbf{q},i\omega_n)=T\sum_{i\nu,i\nu^{\prime}}
   \sum_{{\alpha_{1}\alpha_{2}\atop\alpha_{3}\alpha_{4}}}
   (\mu_J^z)_{\alpha_1\alpha_3}(\mu_J^z)_{\alpha_2\alpha_4} 
   \cdot \chi_{{\alpha_1,\alpha_2\atop
    \alpha_3,\alpha_4}}(i\nu,i\nu^{\prime})_{\textbf{q},i\omega_n}.
\label{eq:chi}
\end{equation}
$\chi(i\nu,i\nu^{\prime})_{\textbf{q},i\omega}$ can be computed by the
Bethe-Salpeter equation using the local irreducible vertex function
$\Gamma_{loc}^{irr}$ (Fig. S2): 
\begin{equation}
\chi_{{\alpha_1,\alpha_2\atop
    \alpha_3,\alpha_4}}(i\nu,i\nu^{\prime})_{\textbf{q},i\omega}=
    [(\chi^{0})_{\textbf{q},i\omega}^{-1}-T\cdot\Gamma_{loc}^{irr}]^{-1}.
\label{eq:BSE_imag}
\end{equation}

\begin{figure}[!h]
\includegraphics[width = \columnwidth]{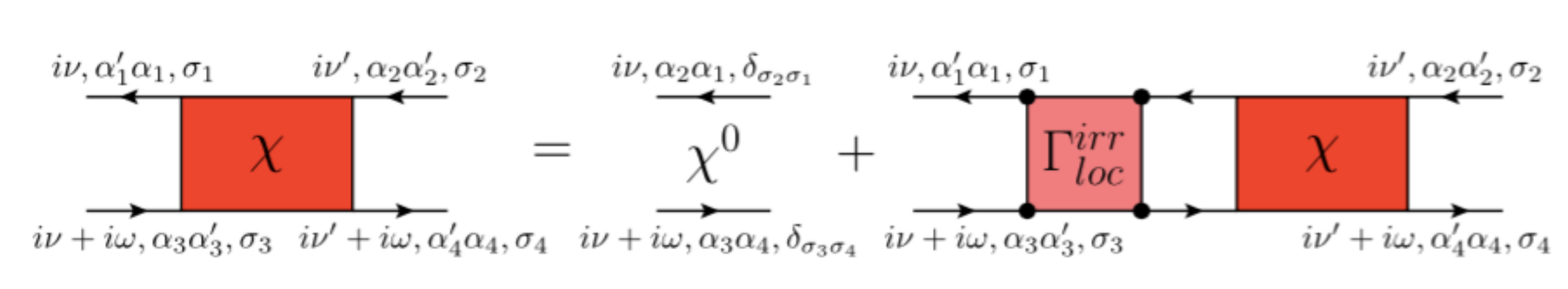}
\caption{The Feynman diagrams for the Bethe-Salpeter equation taken from 
Ref. \citenum{Park:2011fg}. It relates the two-particle GreenÕs function 
with the polarization and the local irreducible vertex function ($\Gamma_{loc}^{irr}$). 
The nonlocal two-particle GreenÕs function is obtained by replacing the local propagator 
by the nonlocal propagator.
\label{FigureS2}}
\end{figure}

Here, the polarization function $\chi^0$ is given as the convolution of the
fully interacting one-particle Green's function $G_{\mathbf{k}}$:
$\chi^0_{\mathbf{q}}=\frac{1}{N_k}\sum_{\mathbf{k}}G_{\mathbf{k}}\cdot 
G_{\mathbf{k+q}}$, and $\Gamma_{loc}^{irr}$ is obtained by the inverse of the 
Bethe-Salpeter equation:
\begin{equation}
\Gamma_{loc{\alpha_1,\alpha_2\atop
    \alpha_3,\alpha_4}}^{irr}(i\nu,i\nu^{\prime})_{i\omega}=
    \frac{1}{T}[(\chi_{loc}^{0})_{i\omega}^{-1}-\chi_{loc}^{-1}]
\end{equation}
where the local two-particle Green's function $\chi_{loc}$ is sampled by the
CTQMC simulations. Finally, the obtained $\chi(\mathbf{q},i\omega_n)$ is
analytically continued ($i\omega_n\rightarrow \omega$) to the real frequency
$\chi(\mathbf{q},\omega)$ using the maximum entropy method.

\begin{figure}[!t]
\includegraphics[width=\columnwidth]{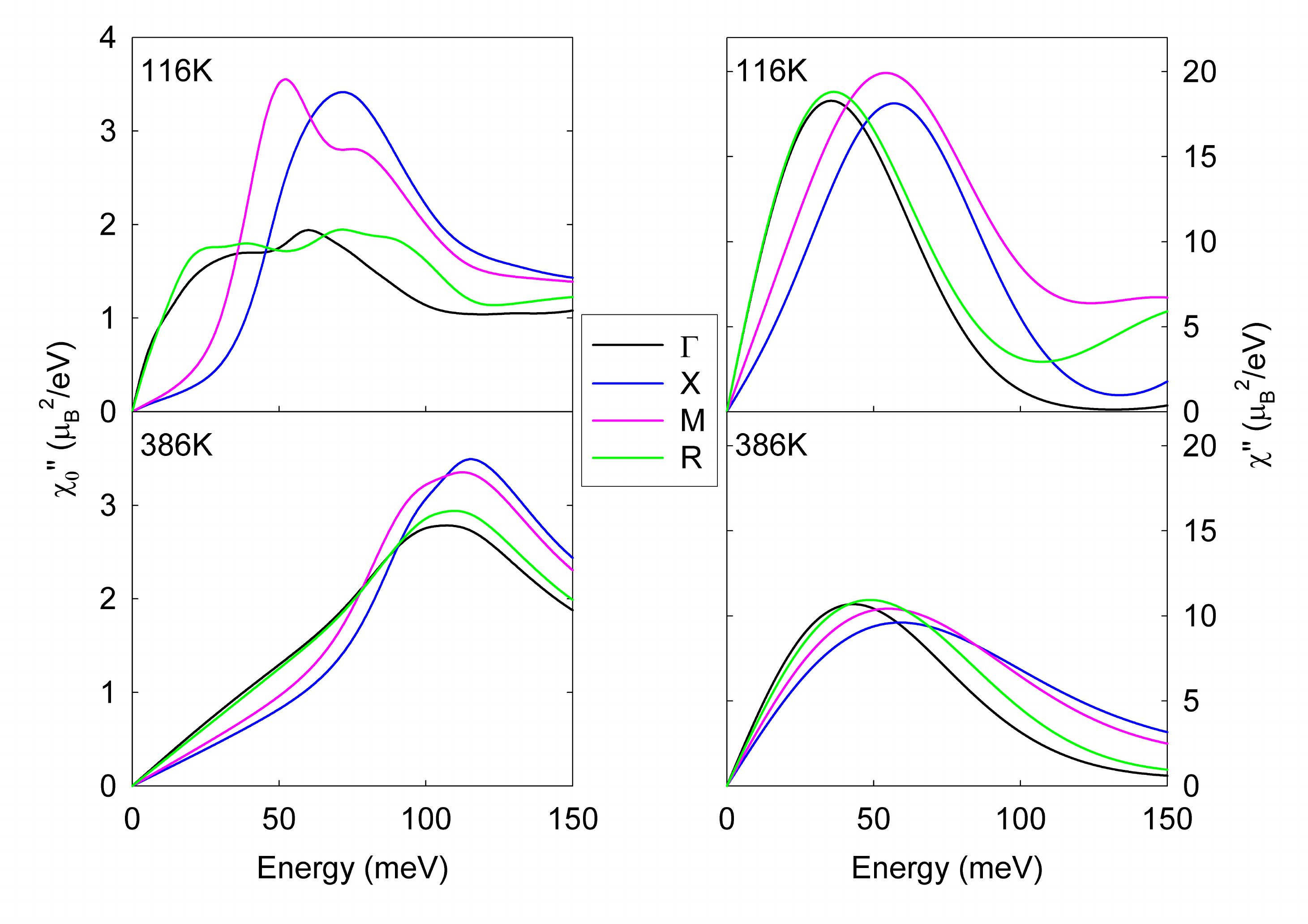}
\caption{Calculated susceptibility at four locations in the Brillouin zone using the 
DFT+DMFT method. The upper and lower panels show calculations at 116~K and 
386~K, respectively (left) before and (right) after applying the two-particle vertex 
correction, $\Gamma_{loc}^{irr}$. 
\label{FigureS3}}
\end{figure}

\begin{figure}[!b]
\includegraphics[width=\columnwidth]{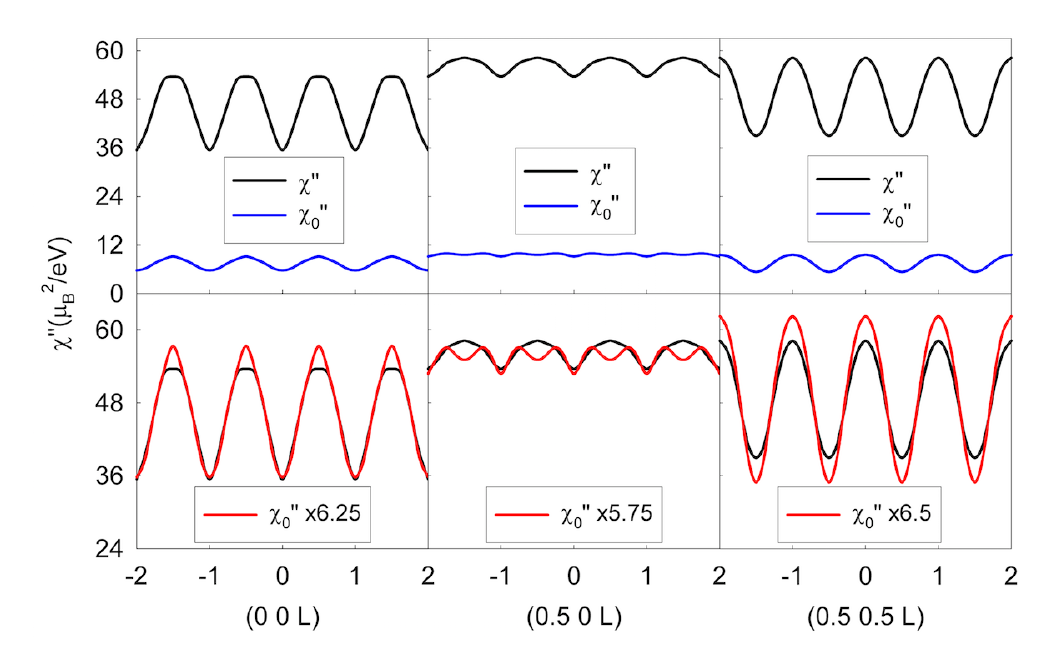}
\caption{Calculated susceptibility along three different directions in reciprocal space
at an energy transfer of 60~meV.  The upper panels show calculations (blue) 
before and (black) after applying the two-particle vertex correction. The lower panels 
shows the same calculations after scaling the uncorrected calculations by 
$\sim$6. 
\label{FigureS4}}
\end{figure}

\subsection{Effect of Two-Particle Vertex Correction}
In Figure S3, we compare the calculated dynamic magnetic susceptibility before and
after including the two-particle vertex correction. Note that this still includes the 
one-particle vertex correction that generates the quasiparticle broadening seen in 
Fig. 1 of the main article. However, the additional correction has two main 
consequences: firstly, some of the fine structure that reflects details of the coherent 
quasiparticle joint densities-of-state is no longer evident, resulting in much smoother
spectra as a function of energy transfer, although the overall energy scales are 
similar. 

Secondly, the susceptibility is strongly enhanced by about a factor 6. This
is most clearly seen in Fig. S4, where the calculations with and without the 
two-particle correction are directly compared. This shows that the main effect of
the correction is to amplify the signal while preserving the Q-dependence seen
in the uncorrected calculations.

\begin{figure}[!b]
\includegraphics[width=0.7\columnwidth]{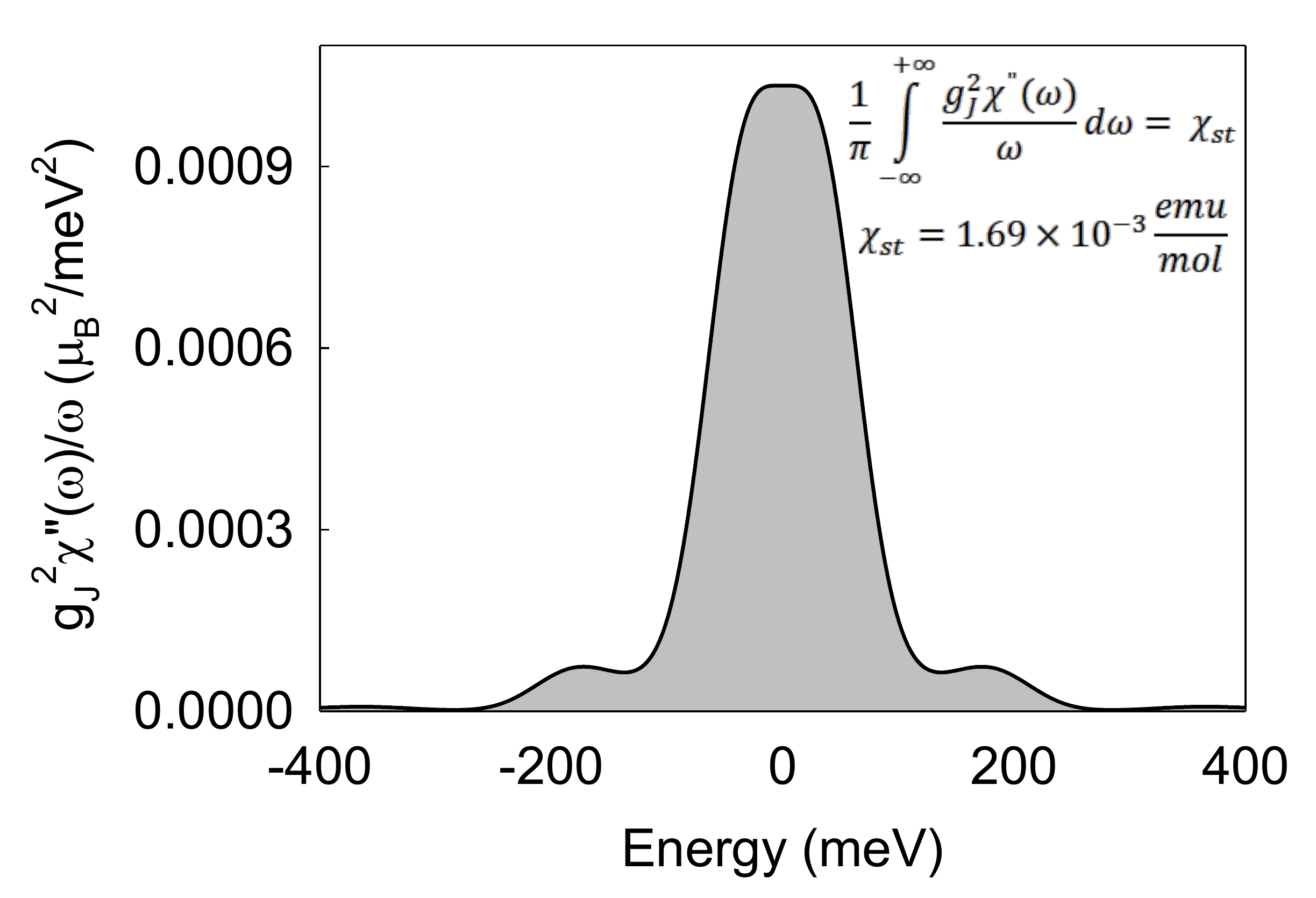}
\caption{Energy dependence of the calculated local susceptibility, \textit{i.e.},
integrated over the first Brillouin zone.  
\label{FigureS5}}
\end{figure}

\section{Experiment}

\subsection{Normalization}
The DMFT+DFT calculations generate $\chi_{zz}(\mathbf{Q},\omega)$ in units of
$\mu_B^2$/meV. For a cubic system such as CePd$_3$, it is possible to derive the 
instantaneous $f$-electron moment, using the moment sum rule \cite{SBalcar:1989us}:
\begin{equation}
\mu_{f}^2= 3g_J^2\mu_B^2\int_{\mathbf{Q}\in BZ}\int_{\omega=-\infty}^{\infty}
    \frac{\chi_{zz}^{\prime\prime}(\mathbf{Q},\omega)}{(1-e^{-\omega/kT})}
    d\mathbf{Q}d\omega
\end{equation}

The calculations, when integrated up to 20~eV, yield an instantaneous moment 
of 6.675~$\mu_B^2$ at 5~K, slightly higher than the expected Ce$^{3+}$ moment of 
6.429~$\mu_B^2$ by about 4\%. Similarly, the static susceptibility can be derived 
using the Kramers-Kronig relation:
\begin{equation}
\chi_{st}=\frac{1}{\pi}\int_{-\infty}^{\infty}{\frac{3g_J^2\chi_{zz}^{\prime\prime}
    (\mathbf{Q}\rightarrow 0, \omega)}{\omega}d\omega}
\end{equation}

This gives $\chi_{st}=1.69\times10^{-3}$~emu/mol, which is also slightly 
higher than the experimental value of $1.4\times10^{-3}$~emu/mol 
\cite{SKappler:1982hw} by about 20\%. These discrepancies are within the 
experimental uncertainties in our comparisons with the neutron data, as 
discussed below. 

The experimental data measured on the MERLIN spectrometer at the ISIS Pulsed
Neutron Source is in the form of $S(\mathbf{Q},\omega)$, which is placed on
an absolute scale after calibration with a standard vanadium sample, using
the standard data reduction procedures implemented by the Horace and Mantid
software packages \cite{SEwings:2016ht,Arnold:2014iy}. The units shown in
figures in the main article can be converted to the calculated units using
\cite{SXu:2013dg}: 
\begin{equation}
\begin{aligned}
S(\mathbf{Q},\omega) [mb/meV sr f.u.] \times 13.77 = \\
    2 g_J^2 f^2(\mathbf{Q}) \chi_{zz}^{\prime\prime}(\mathbf{Q}, \omega) 
    [\mu_B^2/eV f.u.]
\end{aligned}
\end{equation}
ignoring the temperature factor, which is approximately 1 for all our 
measurements. The factor 2 (instead of 3) is because inelastic neutron 
scattering only measures the two components of the dynamic magnetic 
susceptibility that are orthogonal to the experimental wavevector (see 
equation 8 in Ref. \citenum{SXu:2013dg}). 

In our comparisons of the DFT+DMFT calculations with the experimental data,
we corrected the calculations for the Ce$^{3+}$ form factor using the numerical
approximation scheme defined in Ref. \citenum{SAnderson:2006xx}, as 
implemented by the Python package, \textit{periodictable} 
(http://periodictable.readthedocs.io/). Because tabulated values for Ce$^{3+}$ 
are not available, the form factor was approximated by the Pr$^{3+}$ form
factor.

Equation 6 shows that we should be able to convert the DFT+DMFT model
into the neutron cross section by dividing by $13.77/(2 g_J)^2$, \textit{i.e.},
a scale factor of 9.37. As shown in the next section, the fitted value 
was $~\sim$10. 

There is an additional uncertainty due to self-shielding. We are unable to
do a reliable quantitative correction for this, but the single crystal
was approximately 5~mm thick, which would produce a self-shielding 
factor of $\sim$0.82 when perpendicular to the beam, and less at other
rotation angles. This would reduce the scale factor from $\sim$10 to 8 or less.
Because of this and  uncertainties in the theoretical normalization, we
estimate that there is an uncertainty in the quantitative agreement between 
theory and experiment of $\sim$20\%.

\subsection{Backgrounds}
The backgrounds in the inelastic neutron scattering data arose primarily from
single phonon scattering from the sample and the sample environment, which
increases quadratically with wavevector transfer, \textbf{Q}, although there 
were also contributions from multiple scattering. Most of the single phonon 
scattering occurs below an energy transfer of 25~meV, and the backgrounds 
are particularly low above 50~meV (see Fig. S6).

\begin{figure}[!t]
\includegraphics[width=\columnwidth]{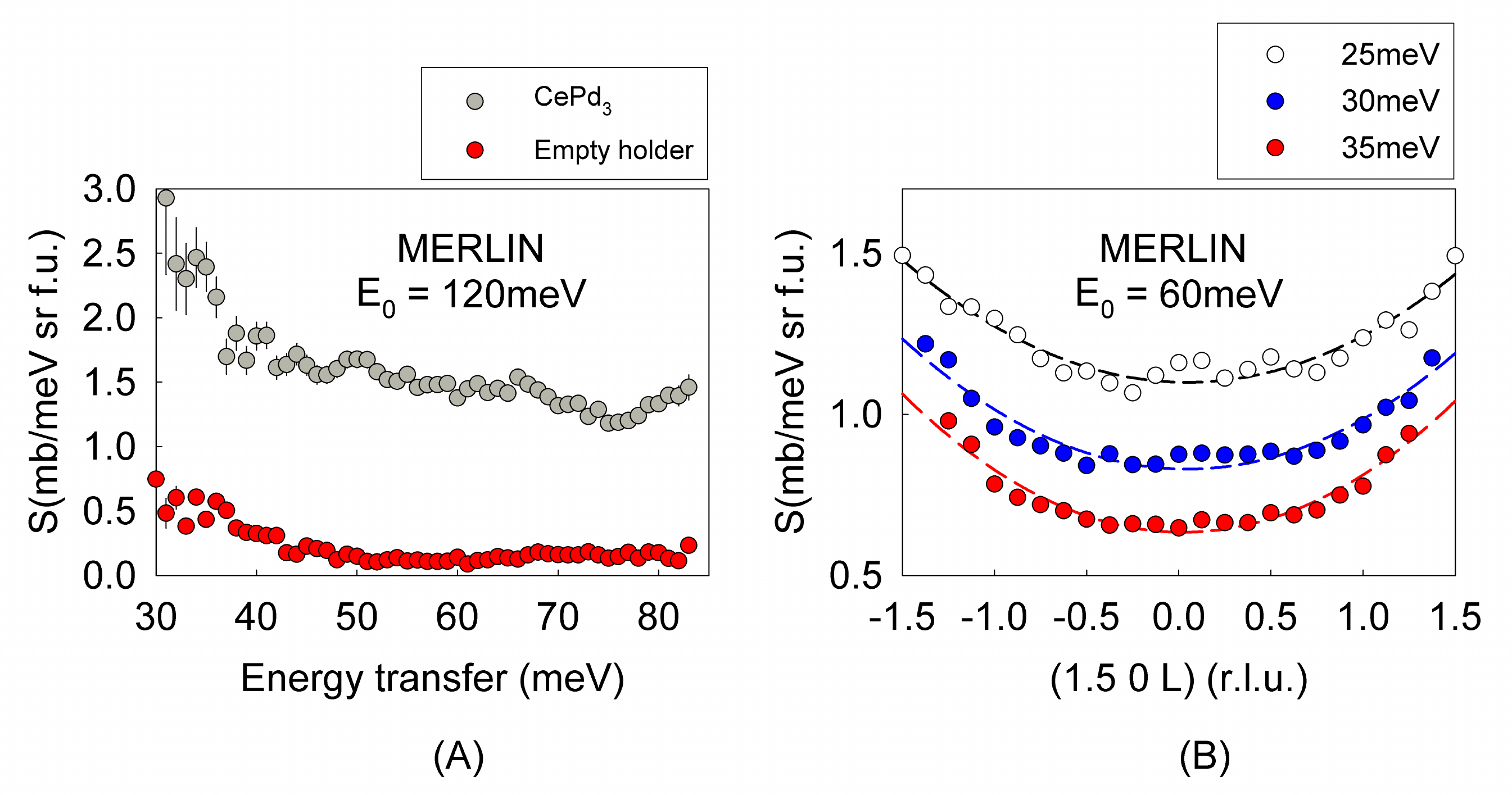}
\caption{(A) Comparison of scattering from CePd$_3$ (grey circles) with the
empty sample holder (red circles), measured with an incident energy of 120~meV 
at the $\Gamma$ point: H = (2.00$\pm$0.25) K = (0.00$\pm$0.25) L =
(0.00$\pm$0.25)). (B) Examples of evolution of the background scattering for
constant energy cuts measured with E$_0$ = 60~meV. The dashed lines are quadratic
fits to the experimental points.
\label{FigureS6}}
\end{figure}

\begin{figure}[!b]
\includegraphics[width=\columnwidth]{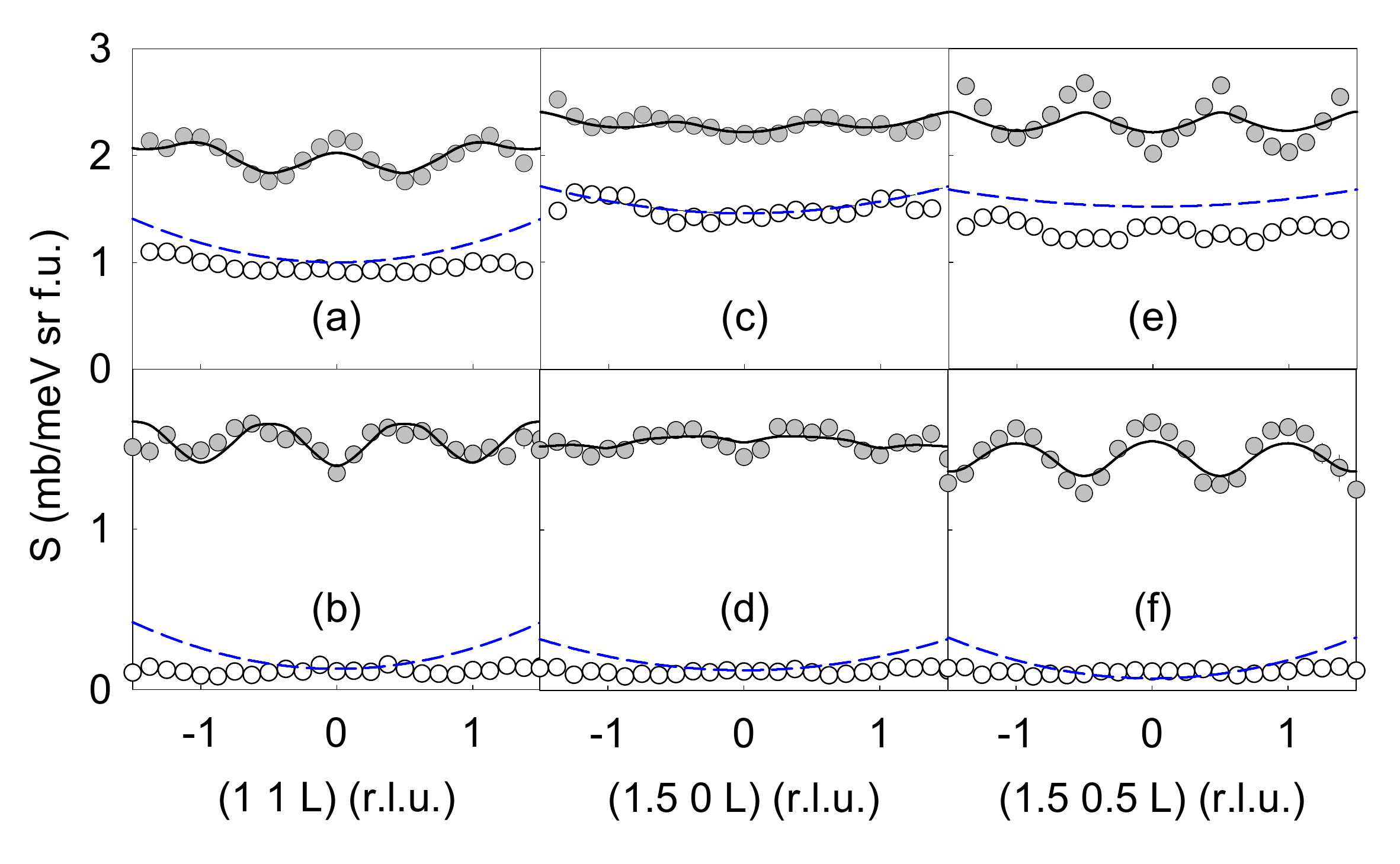}
\caption{Fits of the measured data to the DFT+DMFT model at energy transfers
of 25~meV (upper panels) and 60~meV (lower panels). The data 
(filled circles) were fit to the sum of the model, normalized by a single scale 
factor, and a quadratic background (blue dashed line), giving the black 
solid line. Data with an empty sample holder measured in the same configuration 
are shown as open circles.
\label{FigureS7}}
\end{figure}

\begin{figure}[!t]
\includegraphics[width=\columnwidth]{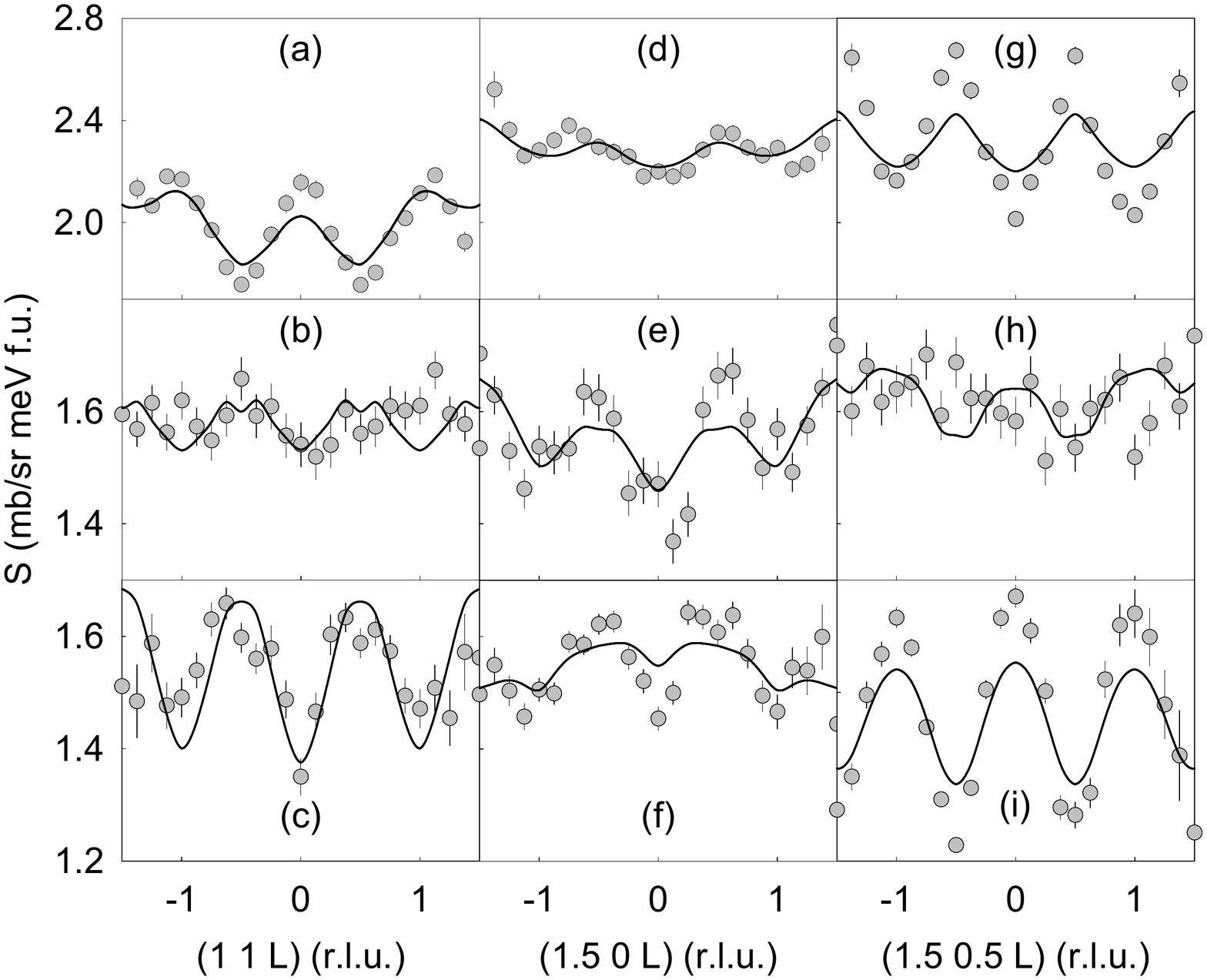}
\caption{One dimensional constant energy cuts along the 
$\Gamma$$\rightarrow$X at H=(1.00$\pm$0.25), K=(1.00$\pm$0.25), 
X$\rightarrow$M at H=(1.50$\pm$0.25), K=(0.00$\pm$0.25) and 
M$\rightarrow$R directions at H=(1.50$\pm$0.25), K=(0.50$\pm$0.25), at energies 
of (a,d,g) 25.0$\pm$2.5~meV, (b,e,h) 45.0$\pm$2.5~meV  and (c,f,i) 60$\pm$5~meV. 
The points represent the inelastic neutron scattering measurements integrated in the 
same energy and H, K ranges.  The lines represent one-dimensional cuts of the 
DFT+DMFT calculations, which have been fit to the data with a single scale factor 
as the only adjustable parameter. The fits include a background that is quadratic 
in Q, which has \textit{not} been subtracted from the 
experimental points (\textit{cf} Fig. 5 in the main article, in which the quadratic
backgrounds have been subtracted).
\label{FigureS8}}
\end{figure}

The scale factor for normalizing the model calculations to the data was therefore
set by fitting the data at an energy transfer of 60 meV to the model plus a 
quadratic background. This gave a value of the scale factor of $\sim$10. As 
the lower panels of Fig. S7 show, the minimum of the estimated quadratic
background agrees very well with the background determined from the empty
sample holder, so the difference between the blue dashed lines and the empty 
circles represents backgrounds from the sample itself. 

The same scale factor was then used in fits of the model+quadratic background
at other energy transfers. The upper panels of Fig. S7 show the fits at $\omega
=25$~meV. Once again, the minima of the estimated backgrounds agree well 
with the empty sample holder measurements, giving confidence that this is a
valid way of estimating the Q-dependent backgrounds.

Fig. S8 shows the same fits as Fig. 5 of the main article, without 
subtracting the backgrounds, to help appreciate the influence of the 
background subtraction on the comparison between theory and
experiment.

\end{document}